\documentclass[conference]{IEEEtran}
\IEEEoverridecommandlockouts

\usepackage{cite}
\usepackage{amsmath,amssymb,amsfonts}
\usepackage{graphicx}
\usepackage{textcomp}
\usepackage[dvipsnames]{xcolor}
\def\BibTeX{{\rm B\kern-.05em{\sc i\kern-.025em b}\kern-.08em
    T\kern-.1667em\lower.7ex\hbox{E}\kern-.125emX}}
\usepackage{graphicx}
\usepackage{subcaption}
\usepackage{booktabs}
\usepackage{multirow}
\graphicspath{ {images/} }
\usepackage[utf8]{inputenc}
\usepackage{fancyhdr}
\usepackage{dirtytalk}
\usepackage{tabularx}
\usepackage[]{nohyperref}
\usepackage{url}
\usepackage[normalem]{ulem} %to strike the words
\usepackage{makecell}
\usepackage{algorithm}
\usepackage{algorithmicx}
\usepackage{algpseudocode}

\usepackage{enumitem,kantlipsum}
\usepackage{ulem}

\usepackage{xcolor}
\usepackage{etoolbox}
\usepackage{tcolorbox}
\tcbuselibrary{breakable, skins}

\definecolor{primaryblue}{HTML}{1B4F9D}
\definecolor{lightbg}{HTML}{EEF3FC}
\tcbset{
  promptstyle/.style={
    enhanced,
    colback=lightbg,
    colframe=primaryblue,
    boxrule=0.8pt,
    arc=6pt,
    left=8pt,
    right=8pt,
    top=6pt,
    bottom=6pt,
    fontupper=\footnotesize,
    coltitle=white,
    fonttitle=\bfseries\itshape,
    colbacktitle=primaryblue,
    attach boxed title to top left={
      xshift=8pt,
      yshift=-2mm
    },
    boxed title style={
      boxrule=0pt,
      arc=6pt,
      left=8pt,
      right=8pt,
      top=4pt,
      bottom=4pt
    }
  }
}

\usepackage{makecell}

\usepackage{caption}
\usepackage{graphicx}
\graphicspath{ {images/} }
\usepackage[utf8]{inputenc}
\usepackage{fancyhdr}
\usepackage{dirtytalk}
\usepackage{tikz}
\usetikzlibrary{shapes.geometric, arrows}
\usepackage{pgfplots}
\pgfplotsset{width=8cm,compat=1.16}
\usepackage[nolist]{acronym}
\begin{acronym}
\acro{LLM}{large language model}
\acro{ML}{machine learning}
\acro{OOD}{out-of-distribution}
\acro{QnA}{questions and answering}
\acro{AI}{artificial intelligence}
\acro{SNAP}{Stanford Large Network Dataset Collection}
\acro{BA}{Barabási–Albert}
\acro{CoT}{chain of thought}

\end{acronym}

\tikzstyle{startstop} = [rectangle, rounded corners, minimum width=2cm, minimum height=0.5cm,text centered, draw=black, fill=red!30]
\tikzstyle{process} = [rectangle, minimum width=2cm, minimum height=0.5cm, text centered, draw=black, fill=orange!30, align=left]
\tikzstyle{decision} = [diamond, minimum width=1.0cm, minimum height=0.4cm, text centered, draw=black, fill=green!30]
\tikzstyle{arrow} = [thick,->,>=stealth]

\newcolumntype{L}{>{$}l<{$}}

\usepackage[labelformat=simple]{subcaption}

\begin{document}
\captionsetup[figure]{labelfont={bf},labelformat={default},labelsep=period,name={Figure}}

% \title{The Power Of Small Language Models With RAG In Telecom Systems}
% \title{An Oracle for 3GPP Standards: \\Phi2-based RAG System with LoRA Fine-Tuning and Enhanced Long-Context Support}
% \title{LLM-Guided A* Search for Shortest Paths in Non-Geometric Network Graphs}
\title{LLM-Aided A* Search in Non-Geometric \\Network Graphs}

% \title{An Oracle for Telecom Networks:\\Leveraging Fine-Tuned Retrieval-Augmented Generation with Long-Context Support}

\author{
    \IEEEauthorblockN{
    Nouf~Alabbasi$^{1}$,
    Esraa~Ghourab$^{1}$,
    Omar~Alhussein$^{1}$
    }
    \IEEEauthorblockA{$^{1}$KU 6G Research Centre, Department of Computer Science, Khalifa University, Abu Dhabi, UAE\\
    Emails: nouf.alabbasi@ieee.org, esraa.ghourab@ku.ac.ae, omar.alhussein@ku.ac.ae}
}

\maketitle

\begin{abstract}

Finding the shortest path in non-geometric network graphs, where edge weights encode arbitrary metrics such as latency or monetary cost rather than spatial distance, poses a challenge for informed search algorithms. Their efficiency depends on an informative heuristic, typically supplied in spatial domains by geometric distances that have no counterpart on non-geometric graphs.
We propose a \ac{LLM}-aided A* algorithm in which an \ac{LLM} generates intermediate waypoints that guide the A* expansion toward promising graph regions. At the core of the approach are landmark distances, which serve both as an admissible landmark-based (ALT) heuristic for the search and as a compact structural feature that, supplied to the \ac{LLM}, restores the distance-to-destination signal it would otherwise lack on non-geometric graphs. Our comprehensive experiments on multiple graph topologies with up to 2,000 nodes demonstrate that \ac{LLM}-generated waypoints reduce the number of expanded nodes by around $50\%$ while incurring only a marginal path cost increase compared to the optimal solution. We further analyze the impact of prompt engineering and show that incorporating compact structural features, namely heuristic estimates, is more effective than advanced prompting techniques. These findings demonstrate the potential of combining \ac{LLM}-based guidance with classical search algorithms for efficient network optimization. Source code and expanded results are available at \href{https://github.com/Nouf-Alabbasi/LLM-aided-A-star-for-networks}{https://github.com/Nouf-Alabbasi/LLM-aided-A-star-for-networks}.

\end{abstract}

\acresetall

\begin{IEEEkeywords}
Network optimization, large language models, LLM-augmented optimization, shortest path routing, A* search
\end{IEEEkeywords}

% ============================ Introduction ============================
% networking context
The rapid evolution of network infrastructure toward 5G and beyond has introduced unprecedented complexity in network operations. Modern architectures built on network function virtualization and software-defined networking shift essential network functions from dedicated hardware to virtual instances, enabling tailored performance guarantees and efficient resource utilization, but at the price of substantially more involved orchestration, resource allocation, and lifecycle management~\cite{luo2025aireasoningwirelesscommunications}. Many of the resulting operational problems, including service chain routing, resource allocation, and traffic engineering, ultimately reduce to graph optimization problems with shortest path computation at their core. Crucially, the underlying network graphs are \emph{non-geometric}, where edge weights can encode arbitrary operational metrics such as latency, monetary cost, or load rather than spatial distance, and nodes possess no meaningful coordinates.

% the problem: geometry-free search
This seemingly innocuous property undermines the workhorse of informed search. The efficiency of A* hinges on an informative admissible heuristic, and in geometric settings, such as road maps and grid worlds, such heuristics come essentially for free in the form of Euclidean or Manhattan distances. On non-geometric network graphs, no such spatial signal exists, and A* degenerates toward uninformed, Dijkstra-like exploration precisely in large complex topologies where efficiency matters most.

% LLMs: direct solving fails; augmentation is promising; but prior work assumes geometry
\begin{figure*}[htbp]
    \centering
    \includegraphics
    [width=\linewidth]{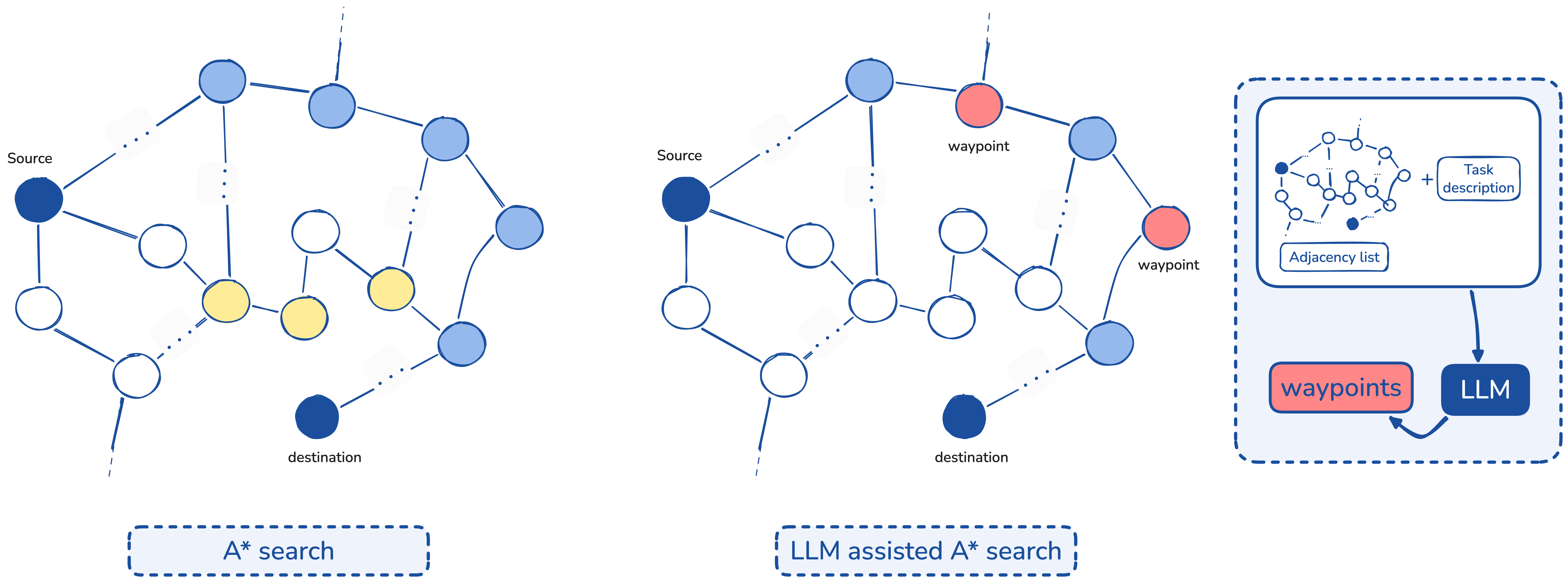}
    \caption{Effect of waypoint guidance on A* exploration. Standard A* (left) is compared with \ac{LLM}-aided A* (middle). Source and destination nodes are shown in dark blue, expanded nodes in yellow, LLM-derived waypoints in pink and the selected path in light blue.}
    \label{fig:arch}
\end{figure*}
In parallel, \acp{LLM} are increasingly being explored for optimization tasks. However, employing them as direct solvers for shortest path problems has so far proven unreliable. \acp{LLM} struggle with multi-step algorithmic reasoning, scale poorly to larger graphs, and are sensitive to how the graph is represented in the prompt~\cite{taylor2024large,liu2023evaluating}. We believe that a more promising direction integrates \acp{LLM} with classical solvers, pairing the model's contextual reasoning and prior knowledge with the algorithmic guarantees of the solver. In particular, recent works in robotics prompt an \ac{LLM} to generate intermediate waypoints that steer A* search on 2D grid maps~\cite{zeng20251000,LLM_A_star_meng2024llm}. These approaches succeed in part because grid environments silently supply two ingredients, namely (i) an admissible geometric heuristic that guides the solver, and (ii) a compact, coordinate-based representation from which the \ac{LLM} can infer proximity and direction. Non-geometric network graphs remove both ingredients. The solver is left without an admissible distance estimate and the \ac{LLM} is left without coordinates from which to form spatial intuition, while the only faithful graph description available, the adjacency list, grows with the size of the graph and strains the prompt.

% our approach + key insight
In this paper, we propose an \ac{LLM}-aided A* algorithm for shortest path computation on non-geometric network graphs that restores both missing ingredients through the use of landmark distances. 
On the solver side, we adopt the landmark-based (ALT) heuristic~\cite{landmark_goldberg2005computing}, which replaces geometric estimates with admissible lower bounds derived from precomputed shortest path distances to a small set of landmark nodes via the triangle inequality. On the \ac{LLM} side, we inject the resulting heuristic estimates into the waypoint generation prompt, where they act as a surrogate coordinate system that restores to the model the distance-to-destination signal it loses on abstract graphs. Guided by this signal, the \ac{LLM} proposes a set of intermediate waypoints, and the estimated distance to the current waypoint augments the A* evaluation function, biasing expansion toward promising regions of the graph while A* remains the underlying solver (Fig.~\ref{fig:arch}).

% contributions as findings
The main contributions of this work are as follows.
\begin{itemize}[leftmargin=*]
    \item We design, to the best of our knowledge, the first \ac{LLM}-guided A* search for shortest paths on non-geometric network graphs, in which landmark distances play a dual role being an admissible heuristic for the solver and a compact structural feature input for the \ac{LLM}.
    \item We show that prompt \emph{content} outweighs prompt \emph{technique}, i.e. enriching the prompt with heuristic estimates raises exploration reduction from $15\%$ to $49\%$, whereas advanced prompting strategies such as \ac{CoT} and few-shot learning substantially help only when the estimates are absent, effectively compensating for missing structural information, and plateau once it is provided.
    % \item We show that the waypoint budget must be calibrated to the expected path length, as over-constraining the search with excessive waypoints diminishes performance gains.
    \item We consider SNAP road network subgraphs and \ac{BA} scale-free graphs with up to 2,000 nodes. Our approach expands around $50\%$ fewer nodes than standard A* while incurring only a marginal average path cost increase of 0.34--0.66, whereas greedy best-first search attains comparable exploration reduction at a path cost increase of 3.92. Moreover, because the \ac{LLM} biases search priorities rather than dictating the solution, we observe empirically that poorly generated waypoints inflate exploration overhead rather than degrade final path quality.
\end{itemize}

% scoping
As a first step, we deliberately study unweighted graphs in order to isolate the effect of \ac{LLM} guidance from constraint handling. Extending the framework to weighted, resource-constrained settings is the subject of ongoing work. It will be considered in an extended version of this paper.
% such as service function chain routing, where the \ac{LLM}'s ability to ingest operator intent and policies expressed in natural language becomes a further asset, is the subject of ongoing work.

The remainder of the paper is organized as follows. Section~\ref{section_related_works} provides an overview of related works. Section~\ref{section_system_model} discusses the proposed \ac{LLM}-aided A* approach. Section~\ref{section_simulations} provides our comprehensive analysis of our approach, and Section~\ref{section_conclusions} concludes the paper and discusses some future research directions.

\section{Related Works}\label{section_related_works}
% In many optimization settings, operator preferences and policies are difficult to explicitly formalize as constraints or objective functions, and conventional methods often require domain-specific data and careful manual design~\cite{wang2017mini}.
% With the rise of \acp{LLM}, their incorporation into optimization is being explored. These models' zero-shot text and code generation capabilities can reduce reliance on extensive datasets and manual engineering while aligning the solver to the operator's preferences. Furthermore, \acp{LLM} can leverage unstructured and heterogeneous data sources enriching the information available to the optimization process beyond what structured inputs alone can provide.
% LLMs as solvers
A growing body of work evaluates \acp{LLM} as direct solvers for graph reasoning problems~\cite{taylor2024large,ye2024language,liu2023evaluating}.
% ~\cite{guo2023gpt4graph,liu2023evaluating,taylor2024large,chen2024exploring,luo2024graphinstruct,zhang2024llm4dyg,ouyang2024gundam,chen2024graphwiz,ye2024language,chai2025graphllm,li2025graph,ge2025can}.
% 
These works analyze the performance of \acp{LLM} on various graph reasoning tasks such as Hamiltonian path and cycle detection, graph connectivity, bipartite matching, topological sorting, maximum flow, and most relevant to this work, shortest path computation.
Several early evaluations find that \acp{LLM} can handle basic structural queries, such as node degree detection and edge existence with reasonable accuracy. However, performance degrades substantially on tasks requiring multi-step algorithmic reasoning, especially on larger graphs \cite{liu2023evaluating,taylor2024large}. In particular, Taylor \textit{et al.} introduced a step-by-step benchmark for classical algorithms including Dijkstra's and BFS, demonstrating that \acp{LLM} struggle to faithfully execute intermediate algorithmic steps even when they occasionally arrive at correct final answers~\cite{taylor2024large}.
Notably, these works are often hindered by the language model's limited reasoning capabilities and are restricted to relatively small graphs with at most 100 nodes.

% LLMs as aid
Rather than relying on \acp{LLM} as standalone solvers, a complementary line of work uses \acp{LLM} to augment classical algorithms. The auxiliary information provided by the \acp{LLM} is used to enhance the outputted solutions and reduce exploration. This augmentation takes several forms.
% tool invokation
Due to the \acp{LLM}' ability to reason over rich contextual descriptions, incorporate domain knowledge, and generalize across problem settings, they are being used to process informal task descriptions and invoke the suitable solver or translate them into formal representations for the solvers~\cite{Zhu_2026_leveraging_LLM_SFC_dep, he2024harnessingexplanationsllmtolminterpreter,LLMAP_2025,API_li2025largelanguagemodelsanalyze}.
Beyond this, \acp{LLM} have been employed to generate algorithm components, such as generating reinforcement learning reward functions, alleviating the burden of manual algorithm design. This approach sidesteps the limitations of \ac{LLM} reasoning by leveraging their code generation capabilities to produce executable programs that solve graph problems algorithmically~\cite{FunSearch_Romera_Paredes_2023,IRL_code_ma2023eureka}.
Finally, in the robotics domain, \acp{LLM} serve as guidance providers to aid in avoiding obstacles along a path. In these approaches, the strengths of the \ac{LLM} in steering exploration complement those of the underlying solver.
Some works use \acp{LLM} to generate intermediate targets along a 2D grid map to aid in pathfinding, while others leverage the \ac{LLM} to suggest plausible exploratory goals for a reinforcement learning agent~\cite{zeng20251000, du2023guidingpretrainingreinforcementlearning, LLM_A_star_meng2024llm}. 

In contrast to these grid-based approaches, which rely on coordinate representations and geometric heuristics, this work targets non-geometric network graphs where neither is available. To the best of the authors' knowledge, this is the first work to integrate \ac{LLM}-generated waypoint guidance with A* search in this setting, using landmark distances in a dual role, as an admissible heuristic for the search and as a compact structural feature in the prompt, and analyzing how such structured inputs influence waypoint quality.

\begin{algorithm}[t]
\caption{A* with LLM Waypoints}
\label{alg:LLM_AStar_in_text}
\begin{algorithmic}[1]

\Require Graph $G=(V,E)$, source $s$, destination $d$, heuristic $h(\cdot,\cdot)$
\Ensure Path $P$ from $s$ to $d$ or \texttt{NoPath}

\State $waypoints \gets \Call{GenerateWaypoints}{LLM, s, d}$% \label{ln:gen_wp}
\State $current\_waypoint \gets waypoints[0]$ %\label{ln:init_wp}

\State Initialize priority queue $Q \gets \emptyset$% \label{ln:init_q}
\State Initialize maps $g(\cdot), parent(\cdot), enqueued(\cdot)$

\State $g(s) \gets 0$ %\label{ln:init\_g}
\State $parent(s) \gets \texttt{None}$
\State $enqueued(s) \gets (0, h(s, d))$
\State \Call{Push}{$Q,(f(s), s)$}

\While{$Q \neq \emptyset$}
    \State $(\_, n) \gets$ \Call{PopMin}{$Q$} %\label{ln:pop}
    \If{$n = d$} %\label{ln:n_d\_check}
        \State $P \gets$ \Call{ReconstructPath}{$parent, d$}
        \State \Return $(P, g(n))$
    \EndIf
    \ForAll{$n' \in Neighbors(n)$} %\label{ln:neighbor\_loop}
        \If{$n' = current\_waypoint \land current\_waypoint \neq d$} 
            \State $current\_waypoint \gets$ $next\_waypoint$
            \State \Call{UpdateQueue}{$Q, current\_waypoint$}
        \EndIf
        % \label{ln:wp\_reached}
        \State $cost \gets 1$
        \State $g_{new} \gets g(n) + cost$
        \If{$n' \in enqueued \land g_{new} \ge g(n')$} %\label{ln:skip}
            \State \textbf{continue}
        \EndIf
        \State $\hat{h}(n') \gets h(n', d) + h(n', current\_waypoint)$ %\label{ln:heuristic}
        \State $g(n') \gets g_{new}$
        \State $parent(n') \gets n$
        \State \Call{Push}{$Q,(g(n')+\hat{h}(n'), n')$} %\label{ln:push}
    \EndFor
\EndWhile

\State \Return \texttt{NoPath}
\end{algorithmic}
\end{algorithm}
% \clearpage

\section{LLM-aided A*}\label{section_system_model}
This section outlines the main elements of our approach. We first provide an overview of the \ac{LLM}-aided search, then expand on the landmark heuristic used in the A* search, followed by an analysis of the waypoint generation prompts. Lastly, we provide an admissibility discussion.

We consider a directed graph $G=(V,E)$ with a source node $s \in V$ and a destination node $d \in V$. For an intermediary node $n \in V$, we denote by $g(n)$ the cost of the best path from $s$ to $n$ found so far, and by $h(n,n')$ a heuristic estimate of the cost from $n$ to a target node $n'$.

\textcolor{black}{The informed search algorithm considered in this work is A*,} a best-first search algorithm that extends Dijkstra's method by integrating a heuristic estimate of the remaining cost to the destination. It evaluates each node $n$ using the evaluation function
\begin{equation}
\label{eq. A* search}
f(n)=g(n)+h(n,d),
\end{equation}
and repeatedly expands the node of lowest $f(n)$. Although A* reduces exploration while maintaining optimality, it relies solely on local heuristic estimates and possesses a limited global view of the graph, which can lead to inefficient search.

Our approach augments A* with a layer of global guidance. Given the adjacency list of $G$ together with a task description specifying $s$ and $d$, we prompt an \ac{LLM} to produce an ordered sequence of intermediate waypoints $\langle \psi_1, \psi_2, \dots, \psi_k \rangle$ that lie along a promising route from $s$ to $d$ (the prompt is detailed in Section~\ref{waypoint_gen}). 
% While the algorithm traverses to the destination, the search also targets
% directed
\textcolor{black}{While the search traverses towards the destination, it targets the intermediate waypoints.}
% The search then targets these waypoints one at a time: 
\textcolor{black}{Our approach} maintains a single \emph{active} waypoint $\psi_i$, and once a node equal to $\psi_i$ is reached, the active waypoint advances to $\psi_{i+1}$. The estimated distance to the active waypoint is added to the A* cost function, so that the evaluation function becomes
\begin{equation}
f(n) = g(n) + h(n,d) + h(n,\psi_i).
\label{eq.cost_func}
\end{equation}
Because the active waypoint $\psi_i$ changes as the search progresses, the second heuristic term retargets accordingly, steering expansion toward the current waypoint and, through the sequence, toward the destination. These waypoints are particularly helpful in regions of the graph with high branching, where they guide the search toward a specific corridor and reduce unnecessary exploration of alternative paths. An illustrative example is shown in Fig.~\ref{fig:arch}.

We now formalize this procedure in Algorithm~\ref{alg:LLM_AStar_in_text}. The algorithm first generates a sequence of waypoints, \textcolor{black}{$\{\psi_1, \psi_2, \dots, \psi_{k}\}$}, and sets the active waypoint to $\psi_1$ (\textit{lines 1--2}), then initializes the priority queue with the source $s$ (\textit{lines 3--8}). The main loop repeatedly dequeues the node $n$ of lowest $f(n)$~\textcolor{black}{(\textit{line 10})}. If $n$ is the destination, the path is reconstructed and returned; otherwise, each neighbor $n'$ of $n$ is evaluated~\textcolor{black}{(\textit{lines 11-15})}. Before computing its priority, the algorithm checks whether $n'$ coincides with the active waypoint and, if so, advances to the next waypoint so that subsequent priorities are computed relative to it~\textcolor{black}{(\textit{lines 16-19})}. The search terminates when the destination is dequeued or the queue is empty~\textcolor{black}{(\textit{line 31})}.

\subsection{A* Heuristic}
The efficiency of A* depends heavily on the quality of its heuristic. Geometric heuristics, such as Manhattan and Euclidean distances, cannot be used on a non-geometric graph, \textcolor{black}{where physical distance is not meaningfully correlated with edge weights.} This is the case for many network graphs, in which edge weights encode arbitrary metrics, such as latency or monetary cost, rather than spatial proximity. Since the optimization objective minimizes these domain-specific metrics, which are not necessarily correlated with physical distance, geometry-based heuristics are inapplicable.

To obtain an admissible heuristic without resorting to a hand-crafted, application-specific design, we adopt the landmark-based (ALT) heuristic~\cite{landmark_goldberg2005computing}. A small set of landmark nodes $L$ is selected, and the shortest-path distances between every node and each landmark are precomputed offline using Dijkstra's algorithm. During search, these precomputed distances yield a lower bound on the distance from a node $n$ to the destination $d$ via the triangle inequality. For any landmark $\ell \in L$,
\begin{equation}
\delta(n,d) \;\ge\; |\, \delta(n,\ell) - \delta(d,\ell) \,|,
\label{eq.triangle}
\end{equation}
where $\delta(\cdot,\cdot)$ denotes the true shortest path distance. Taking the tightest such bound over all landmarks gives the heuristic
\begin{equation}
h(n,d) \;=\; \max_{\ell \in L} \, |\, \delta(n,\ell) - \delta(d,\ell) \,|.
\label{eq.alt}
\end{equation}
By construction, $h(n,d)$ never exceeds the true distance, so the heuristic is admissible while remaining informative. Moreover, because the landmark distances are computed offline, the heuristic is especially well suited to settings where the same graph serves many source--destination queries, as is common in network routing.

\subsection{Waypoint selection}\label{waypoint_gen}
To analyze the effect of prompt engineering on waypoint quality, we construct several waypoint selection prompts. All of them share the common template shown in \textit{Prompt~1}, which specifies the source and destination nodes, a description of the search task, the adjacency list, and tie-breaking instructions; the prompts to follow differ only in additional guidance they provide.

The first prompt contains only the adjacency list, with no additional guidance, and tests whether the \ac{LLM} can reason about the route from topology alone. The second prompt supplements the adjacency list with the heuristic estimates to the destination, the $h$-values. In grid environments, the \ac{LLM} can infer proximity from node coordinates, but such geometric reasoning is unavailable on abstract network graphs. By supplying the $h$-values explicitly, we provide a compact, task-relevant signal that approximates distance-to-destination information, and we test whether this structural guidance improves waypoint quality.

Finally, we examine two reasoning-enhancing techniques, namely \ac{CoT} and few-shot prompting. \ac{CoT} has been shown to improve \ac{LLM} performance on multi-step reasoning tasks; since waypoint selection requires reasoning over connectivity across multiple hops, it is a natural candidate. We test few-shot prompting by including examples of the A* search process in the prompt, allowing us to observe whether in-context demonstrations help the \ac{LLM} internalize what constitutes a useful waypoint. The intuition is that exposing which nodes A* expands, and in what order, may help the \ac{LLM} extract implicit search patterns and thereby improve its waypoint choices.

\begin{tcolorbox}[breakable,colframe=primaryblue,colback=lightbg,title=Prompt 1. Waypoint Generation Prompt Template]
\footnotesize
You are a network optimization assistant tasked with finding the shortest path in a directed graph.\\
\\
PROBLEM SETUP:\\
- Network representation: Directed graph as adjacency list\\
- Input: Start node $\{s\}$, destination node $\{d\}$\\
- Output: Shortest path as ordered list of node IDs\\
\\
AVAILABLE INFORMATION:\\
Adjacency list (graph structure):\\
$\{adjacency\_list\}$\\

CRITICAL STRATEGY - TIE-BREAKING WITH LOOKAHEAD:\\
When choosing the next node, if two or more candidates have the same number of edges traversed:\\
1. Do NOT explore multiple branches\\
2. Instead, look ahead to their immediate neighbors\\
3. For each tied candidate, identify its neighbors\\
4. This lookahead ensures you pick the node that opens the best path forward\\
\\
INSTRUCTIONS:\\
$\{$instruction text$\}$
\\
\\
From the identified path, select a set of $\{num\_waypoints\}$ key waypoints.\\
Answer: [$node\_ID$, $node\_ID$, ..., $\{goal\}$]\\
\end{tcolorbox}

\subsection{Admissibility}
The standard A* algorithm returns an optimal path when its heuristic is admissible, that is, when it never overestimates the true cost to the destination. If admissibility is violated, the search may assign an inflated estimate to a node on the optimal path, causing it to be deprioritized in favor of suboptimal alternatives. Conversely, if the estimate $h(n,d)$ never overestimates the true cost from $n$ to $d$, A* is guaranteed to return the shortest path. The landmark heuristic of Eq.~\eqref{eq.alt} satisfies this condition by construction, since the triangle inequality can only underestimate the true shortest-path distance.

In our approach, however, the evaluation function in Eq.~\eqref{eq.cost_func} adds the waypoint term $h(n,\psi_i)$ to the admissible estimate $h(n,d)$. Because this extra term may cause the combined estimate to overestimate the true cost to the destination, the augmented search is no longer strictly admissible. Nonetheless, our experimental results in Section~\ref{sec:results_prim} show that the substantial reduction in exploration is accompanied by only a marginal increase in path cost, and in many cases no increase at all. While the present work establishes this trade-off empirically, in an extended version of this work we will formally characterize it by deriving a bound on the suboptimality of the returned path, i.e. the worst-case excess over the optimal cost introduced by the waypoint term.

\section{Experimental Results}\label{section_simulations}

\begin{figure}[t]
  \centering
  \begin{subfigure}[b]{0.48\linewidth}
    \centering
    \includegraphics[width=\linewidth]{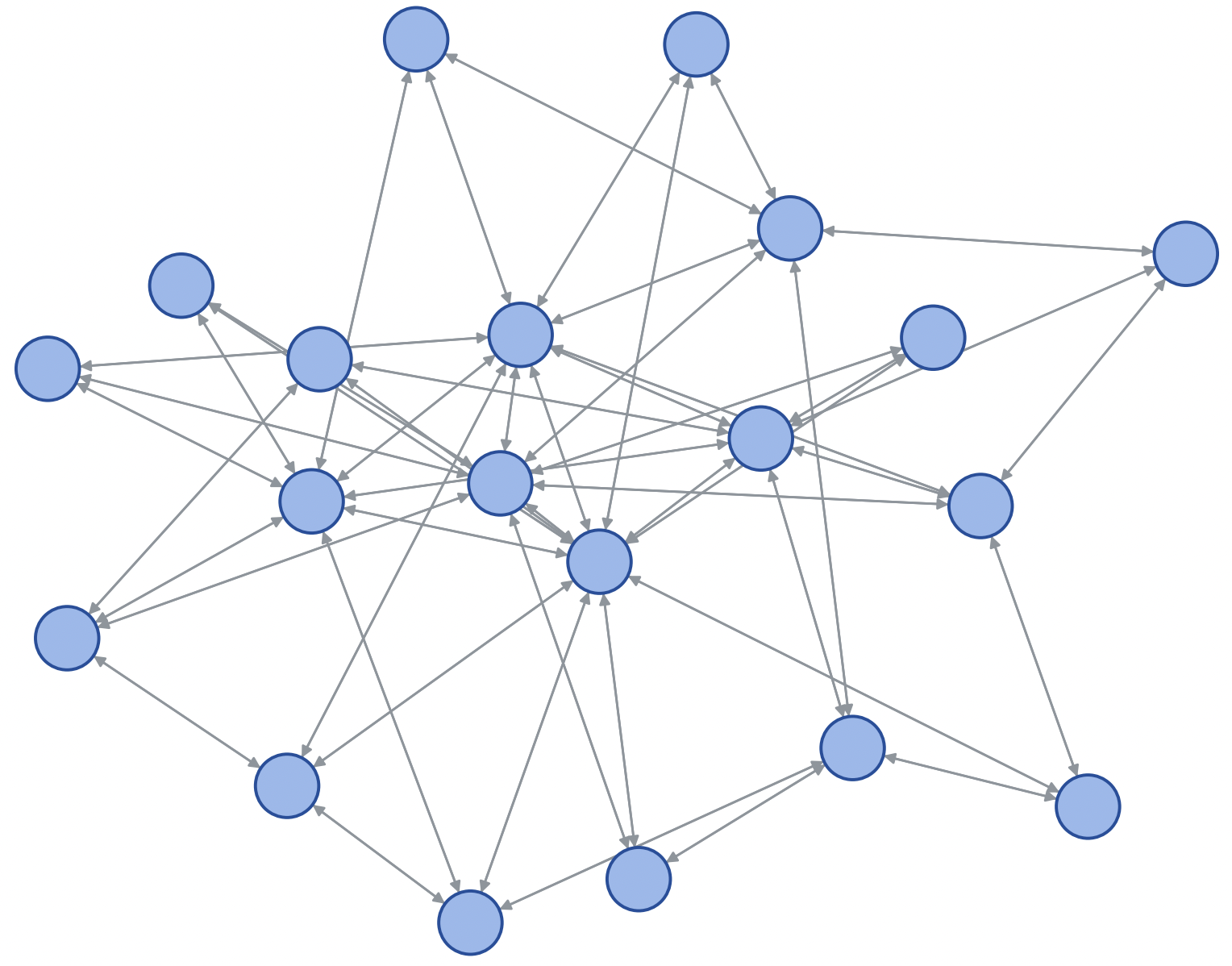}
    \caption{Barabási–Albert model graph}
    \label{fig:BA_scale}
  \end{subfigure}
  % \hfill
  \begin{subfigure}[b]{0.48\linewidth}
    \centering
    \includegraphics[width=\linewidth]{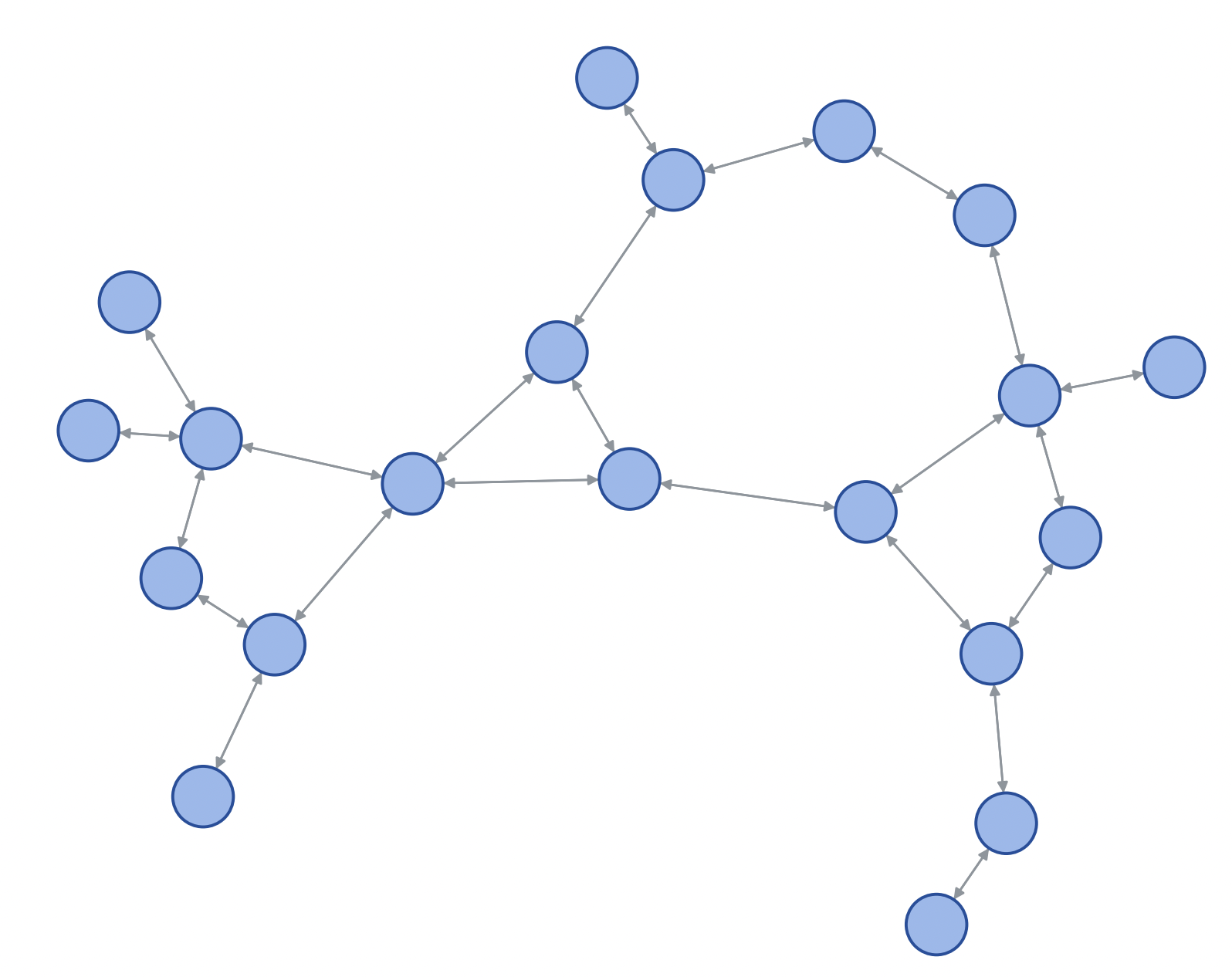}
    \caption{SNAP graph}
    \label{fig:SNAP_fig}
  \end{subfigure}
  \caption{Representative subgraphs: (a) Barabási–Albert (BA) model graph exhibiting a scale-free topology, and (b) SNAP road network graph.}
  \label{fig:graph figures}
\end{figure}

\begin{table}[!t]
    \centering
    \caption{Effect of the number of LLM-derived waypoints on exploration reduction across different SNAP graph sizes. Results report the percentage reduction in expanded nodes achieved by LLM-aided A* for varying waypoint counts (5, 8, and 12) and graph sizes ranging from 750 to 2000 nodes. For each size, the largest reduction is shown in bold.}
    \label{fig:table_2}
    \setlength{\tabcolsep}{6pt}
    \renewcommand{\arraystretch}{1.15}
    \begin{tabular}{ccc}
        \toprule
        \textbf{\# of nodes} & \textbf{\# of waypoints} & \textbf{Exploration reduction (\%)} \\
        \midrule
        \multirow{3}{*}{750}  & 5  & 51.25 \\
                              & 8  & \textbf{53.44} \\
                              & 12 & 45.58 \\
        \midrule
        \multirow{3}{*}{1000} & 5  & 51.64 \\
                              & 8  & \textbf{52.32} \\
                              & 12 & 17.11 \\
        \midrule
        \multirow{3}{*}{1500} & 5  & 51.39 \\
                              & 8  & \textbf{54.42} \\
                              & 12 & 16.16 \\
        \midrule
        \multirow{3}{*}{2000} & 5  & \textbf{49.95} \\
                              & 8  & 28.40 \\
                              & 12 & 37.84 \\
        \bottomrule
    \end{tabular}
\end{table}

\subsection{Experimental Setup}
\label{sec:experimental}
% \textcolor{magenta}{did you mention anywhere that you are dealing with hop-based distance. If not, add this consideration earlier please}
To evaluate the effectiveness of our \ac{LLM}-aided A* approach on network graphs, we conduct a systematic set of experiments across multiple graph instances, source–destination pairs, and prompt configurations.
% graphs
The first type of graph on which we conduct experiments consists of subgraphs from the \ac{SNAP} road network graphs~\cite{leskovec2016snapgeneralpurposenetwork}.
Additionally, we conduct our experiments on the \ac{BA} graph model, which creates scale-free graphs through preferential attachment, as depicted in Fig.~\ref{fig:BA_scale}. These graphs represent two distinct graph topologies, capturing both synthetic scale-free structures and real-world road network characteristics.
% 
% iterations
We conduct experiments on unweighted graphs (where edge costs are uniform hop counts) with 500 to 2,000 nodes. For each size, we generate 5 graph instances with 10 randomly chosen source–destination pairs each, yielding 50 routing queries per size. For each graph, we select 20 landmarks using the farthest-point heuristic. In this work, we measure the performance of our approach as the reduction in expanded nodes. A node is considered expanded when it is 
% to refer to the nodes that are 
popped from A*'s priority queue, and its neighbors are evaluated.
All our experiments use OpenAI's GPT-4.1 as the underlying language model for waypoint generation. This model offers a favorable balance between instruction-following capability and inference efficiency for tasks where the prompt contains large structured inputs.
% model
Notably, GPT-4.1 is not a reasoning model and does not perform multi-stage internal reasoning chains~\cite{openai_gpt41}. In this work, maintaining a predictable inference time is a key factor in ensuring measurable time complexity. Though repeatedly prompting the \ac{LLM} to critique and revise its output could improve waypoint quality, our work prioritizes temporally consistent waypoint generation, making the non-reasoning, single-pass inference of GPT-4.1 well suited for this work\footnote{Source code and expanded results are available at \href{https://github.com/Nouf-Alabbasi/LLM-aided-A-star-for-networks}{https://github.com/Nouf-Alabbasi/LLM-aided-A-star-for-networks}}. Here, we report expansion count as the primary metric, with waypoints generated once per query in an offline-amortizable step; wall-clock comparison is left to future work.

\subsection{Results and Analysis}
\label{sec:results_prim}
 
\begin{figure}
    \centering
    \includegraphics[width=\linewidth]{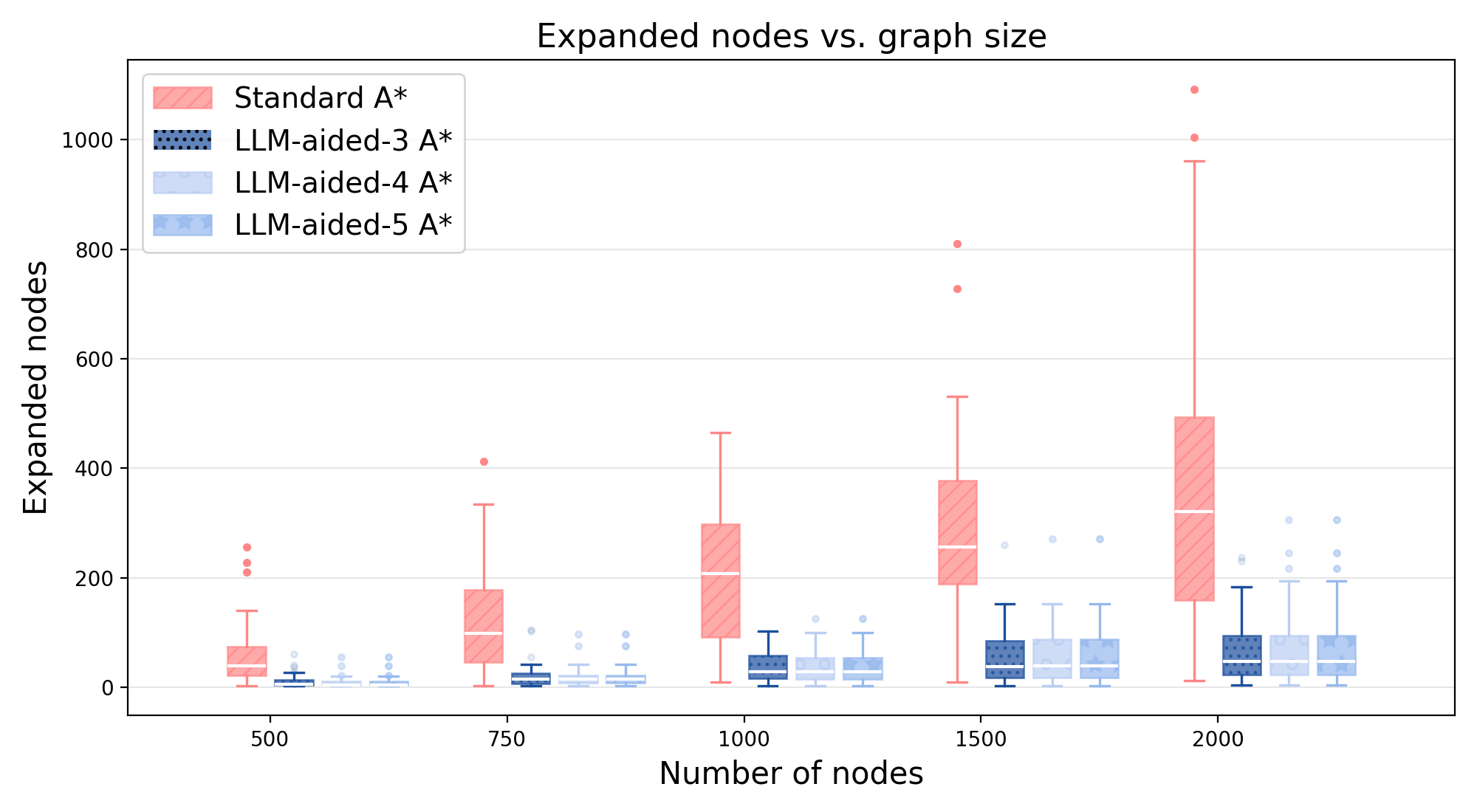}
    \caption{Expanded nodes versus graph size for \ac{BA} scale-free graphs with edge degree 4. Results compare standard A* with LLM-aided A* using 3, 4, and 5 waypoints across graph sizes ranging from 500 to 2000 nodes.}
    \label{fig:num_waypoints_BA}
\end{figure}
 
\begin{figure}
    \centering
    \includegraphics[width=\linewidth]{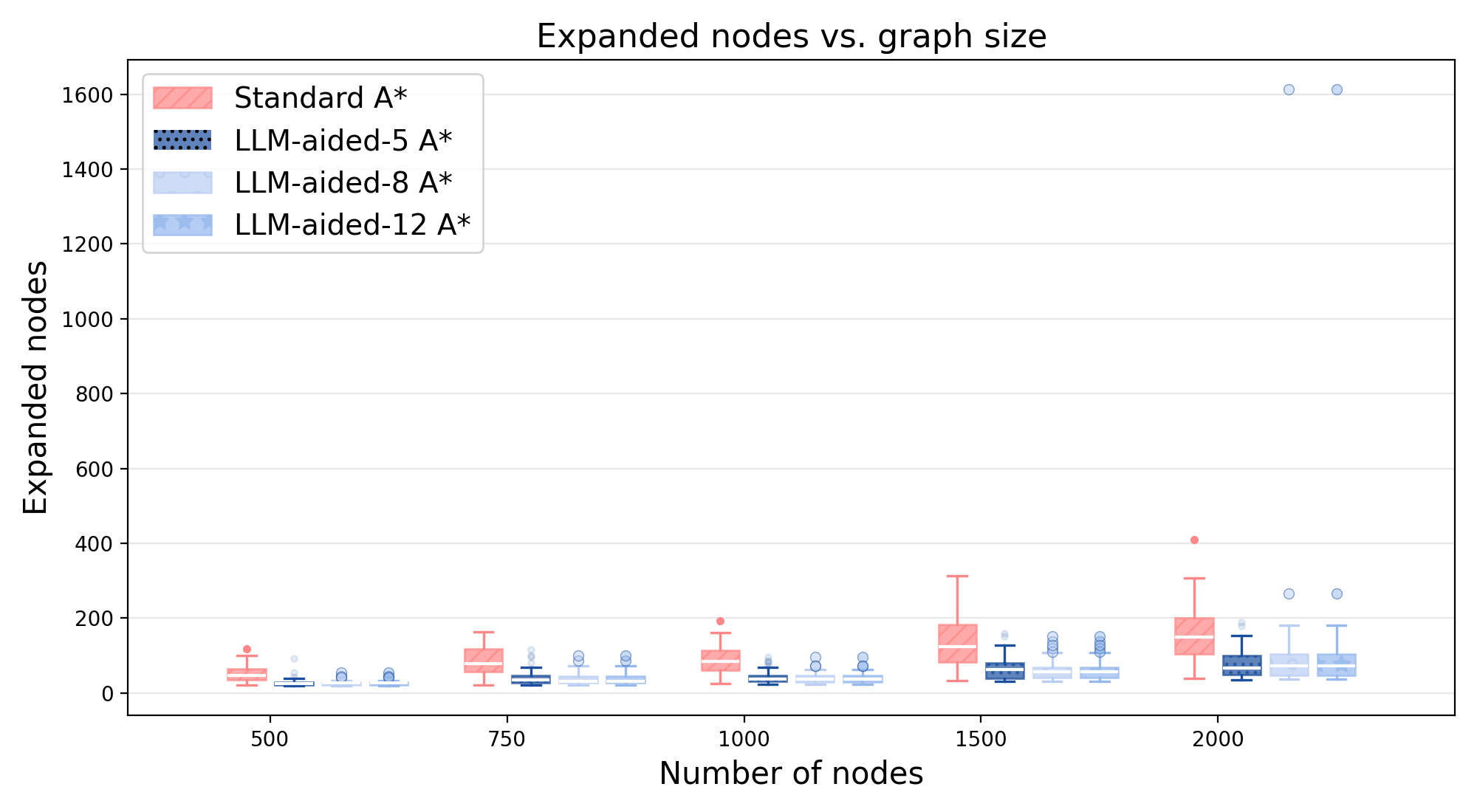}
    \caption{Expanded nodes versus graph size for SNAP road network subgraphs. Results compare standard A* with LLM-aided A* using 5, 8, and 12 waypoints across graph sizes ranging from 500 to 2000 nodes.}
    \label{fig:num_waypoints_road}
\end{figure}

% ============================
% ============================ expermental recap
In this section, we evaluate the effect of waypoint count, then analyze prompting strategies, and finally examine scaling behavior of our approach.
% 
% ============================
% ============================ effect of waypoints Table \ref{table_2} and Figs~\ref{fig:num_waypoints_BA} and~\ref{fig:num_waypoints_road}
    % we adopt 5 waypoints
% \textcolor{magenta}{the following section- does it seem cohesive? is the idea consistent??}
To assess the effect of waypoint count on performance, we evaluate our \ac{LLM} guided A* with different numbers of generated waypoints for each graph size. The results are presented in Table~\ref{fig:table_2} and visualized in Figs.~\ref{fig:num_waypoints_BA} and~\ref{fig:num_waypoints_road}.
In the \ac{SNAP} graphs, all three waypoint counts (5, 8, and 12) reduce the number of expanded nodes compared to standard A*, as seen by the consistently lower medians and tighter interquartile ranges of the \ac{LLM}-aided variants in Fig.~\ref{fig:num_waypoints_road}. 
% A node is considered expanded when it is removed from the priority queue and its outgoing neighbors are generated.
We also notice that larger waypoint counts yield diminishing reductions compared to their lower-count counterparts.
For instance, in Table~\ref{fig:table_2}, it can be seen that leveraging 12 waypoints yields diminishing reductions compared to their lower-count counterparts. 
This behavior arises because, in our approach, the \ac{LLM} generated waypoints are meant to bias the search towards more promising regions. With the right number of waypoints, the search receives a general directional guide while retaining the flexibility to identify the optimal route within those regions.
% \textcolor{orange}{but with an excessive number of waypoints, the search is drawn toward too many intermediate targets, diminishing the influence of the goal-directed heuristic, resulting in mroe exploration + if we provide way too little waypoints, the search isn't guided enough.}
% Similarly, based on the results where the expanded nodes reduction at 500 nodes with 5 waypoints is $50\%$ and the limited guidance provided by only 3 points leads to an expanded nodes reduction of $30\%$ 
The results on \ac{BA} graphs reveal a complementary pattern, as plotted in Fig.~\ref{fig:num_waypoints_BA}. Because these graphs have smaller diameters, we use correspondingly lower waypoint counts (3, 4, and 5), all below the average diameter for each graph size. On smaller graphs, all three counts yield similar reductions. However, as graph size increases, the 5-waypoint variant achieves greater reductions, suggesting that longer paths benefit from additional intermediate targets to maintain directional focus across a larger search space. Taken together, these results suggest that waypoint count should be calibrated to the expected path length. 
It can be seen that LLM-aided A* exhibits significantly improved scaling behavior compared to standard A*, with substantially fewer node expansions as graph size increases. While standard A* shows rapid growth in explored nodes, the LLM-guided variants maintain a slower growth rate, indicating more efficient search.

\begin{table*}[t]
    \centering
    \caption{Exploration reduction achieved by the proposed LLM-aided A* across different prompt configurations for \ac{SNAP} graphs with 1000 nodes. The table reports the percentage reduction in the number of expanded nodes relative to standard A*, along with the average and median number of expanded nodes for both A* and LLM-aided A*. Results are shown for various prompt variations, as well as for a greedy best-first search using $h(n)$. The corresponding average path cost increase is also reported. The best reduction and lowest cost increase are shown in bold.}
    \label{fig:results_1}
    \footnotesize
    \setlength{\tabcolsep}{5pt}
    \renewcommand{\arraystretch}{1.3}
    \begin{tabular}{llcccc}
        \toprule
        \multicolumn{2}{c}{\textbf{Prompt}}
            & \makecell{\textbf{Expanded node}\\\textbf{reduction (\%)}}
            & \makecell{\textbf{Expanded nodes}\\\textbf{(A* $\vert$ LLM-aided A*)}}
            & \makecell{\textbf{Median expanded nodes}\\\textbf{(A* $\vert$ LLM-aided A*)}}
            & \makecell{\textbf{Average}\\\textbf{cost increase}}\\
            % & \makecell{\textbf{Average}\\\textbf{\# waypoints}} \\
        \midrule
        \multirow{4}{*}{$adj\_list$}
            & --        & 15.05 & 76.4 $\vert$ 64.9  & 77 $\vert$ 39   & 0.35\\
            & CoT       & 40.63 & 76.4 $\vert$ 45.36 & 77 $\vert$ 38   & 0.37\\
            & Few-Shot  & 36.13 & 76.4 $\vert$ 48.80 & 77 $\vert$ 39   & 0.44\\
            & No waypoint limit & 27.12 & 76.4 $\vert$ 55.68 & 77 $\vert$ 42.5 & 0.50\\
        \midrule
        \multirow{4}{*}{$adj\_list\_h\_values$}
            & --        & \textbf{48.69} & 76.4 $\vert$ 39.20 & 77 $\vert$ 32.5 & 0.66\\
            & CoT       & 47.93 & 76.4 $\vert$ 39.78 & 77 $\vert$ 36   & 0.56\\
            & Few-Shot  & 46.23 & 76.4 $\vert$ 41.08 & 77 $\vert$ 33   & \textbf{0.34}\\
            & No waypoint limit & 48.35 & 76.4 $\vert$ 39.46 & 77 $\vert$ 31 & 0.48\\
        \midrule
        \multicolumn{2}{l}{Greedy best-first search} & 45.05 & 76.4 $\vert$ 43.50 & 77 $\vert$ 39 & 3.92\\
        \bottomrule
    \end{tabular}
\end{table*}
 
\begin{figure}[t]
    \centering
    \includegraphics[width=\linewidth]{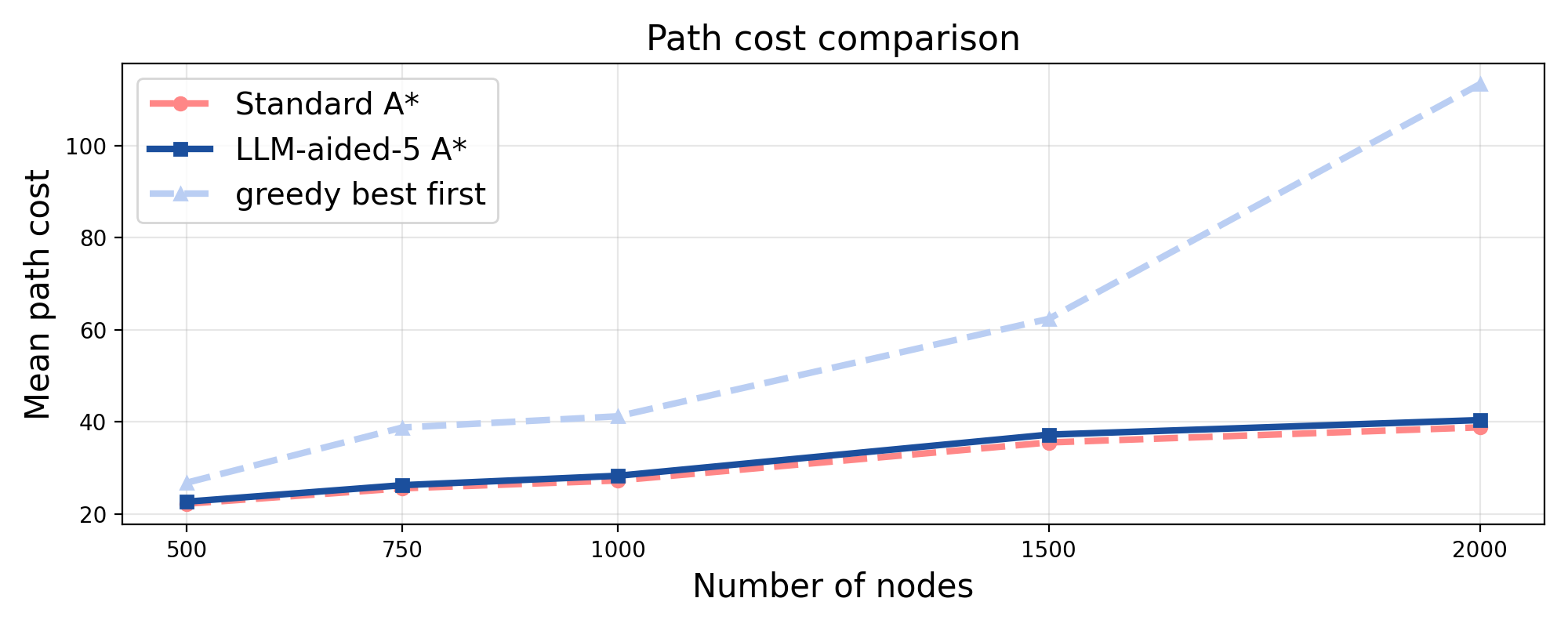}
    \caption{Path cost comparison across varying graph sizes. Results are averaged over 50 runs}
    \label{fig:cost_comparison}
\end{figure}
 
% ============================
% ============================ effect of prompting strategy \ref{results_1), concluding with the h-values insight
The effect of prompt engineering on the exploration reduction can be seen in Table \ref{fig:results_1} for SNAP graphs with 1,000 nodes.
The average and median number of expanded nodes for A* and LLM-aided A* search is also recorded. Additionally, we report the average path cost increase relative to standard A*, along with the average number of waypoints generated. It can be seen that providing $h$-values to the \ac{LLM} dramatically improves exploration reduction. For instance, the base $adj\_list\_h\_values$ (i.e., providing both adjacency list with the $h$-values) prompt achieves~$48.69\%$ reduction versus only~$15.05\%$ for $adj\_list$ alone. Furthermore, prompting techniques such as \ac{CoT} and few-shot enhance the performance considerably when the \ac{LLM} is not provided with the $h$-values, jumping from $15\%$ to $40\%$. However, these improvements plateau when $h$-values are already provided.
This suggests \ac{CoT} and few-shot prompts compensate for the absence of adequate graph structural information. In other words, without the $h$-values, the \ac{LLM} must infer node-to-destination proximity from the adjacency structure alone, which these prompting techniques help facilitate. 
% Claude
When $h$-values are provided directly, this inference becomes unnecessary, and the marginal benefit of advanced prompting diminishes. More broadly, this implies that for graph-based tasks, enriching the prompt with task-relevant structural features is more impactful than sophisticated prompting strategies applied to the graph representation.

% ============================
% ============================ comparison with greedy best-first search - cost vs exlploration trade off - (LLM guides, A* solves, robust to hallucinations, outlier figure)
To contextualize these gains, we compare to greedy best-first search which forgoes optimality guarantees in favor of aggressively reducing the exploration. While it achieves comparable exploration reduction of approximately $45\%$, it incurs a vastly higher cost increase of 3.92, compared to the LLM-aided A* with cost increase in the 0.34–0.66 range. We plot the average of the cost of the path found for each graph size for standard A*, our LLM-aided A* and greedy best first search, as seen in Fig.~\ref{fig:cost_comparison}.
This disparity highlights the key advantage of the proposed approach, the \ac{LLM} provides guidance to the algorithm without significantly degrading path quality.
This is largely due to the nature of our approach where the \ac{LLM} serves only as a guide, while A* remains the underlying solver.
 
\begin{figure*}[t]
\centering
 
% -------- Top row: Standard A* --------
\begin{subfigure}[t]{0.32\textwidth}
\includegraphics[width=\linewidth]{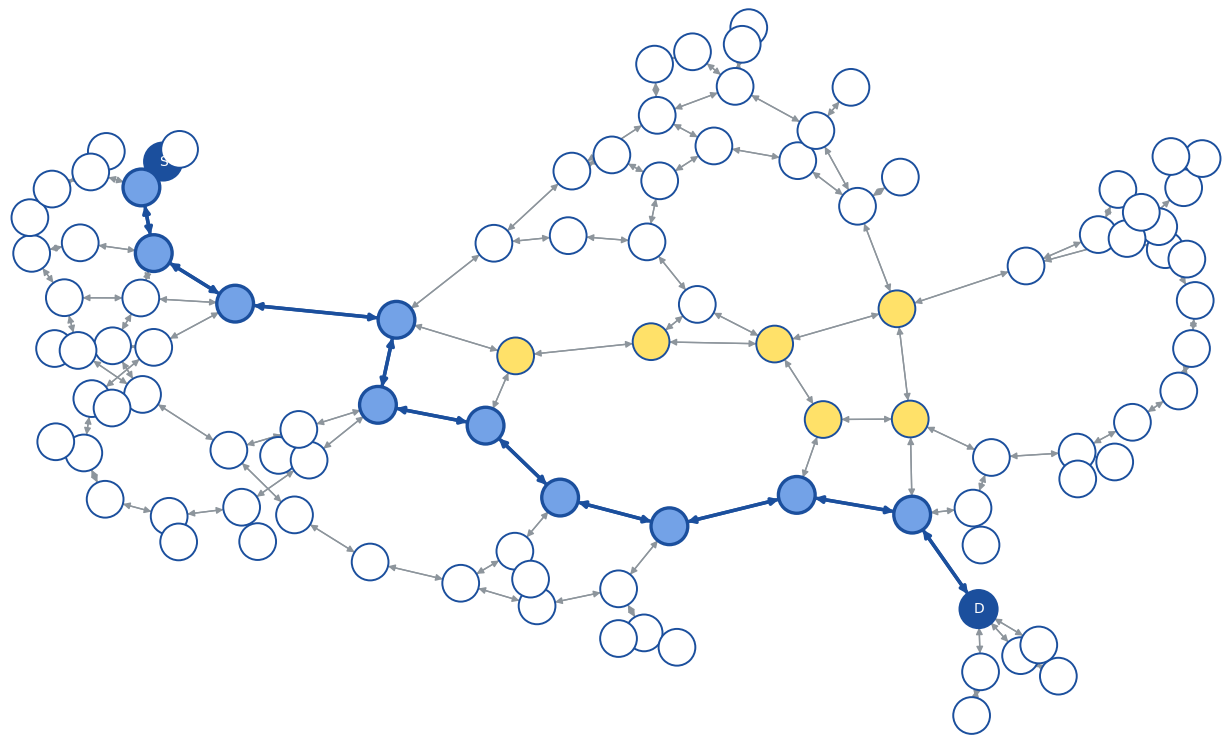}
\caption{Standard A* - 100 nodes}
\end{subfigure}
\hfill
\begin{subfigure}[t]{0.32\textwidth}
\includegraphics[width=\linewidth]{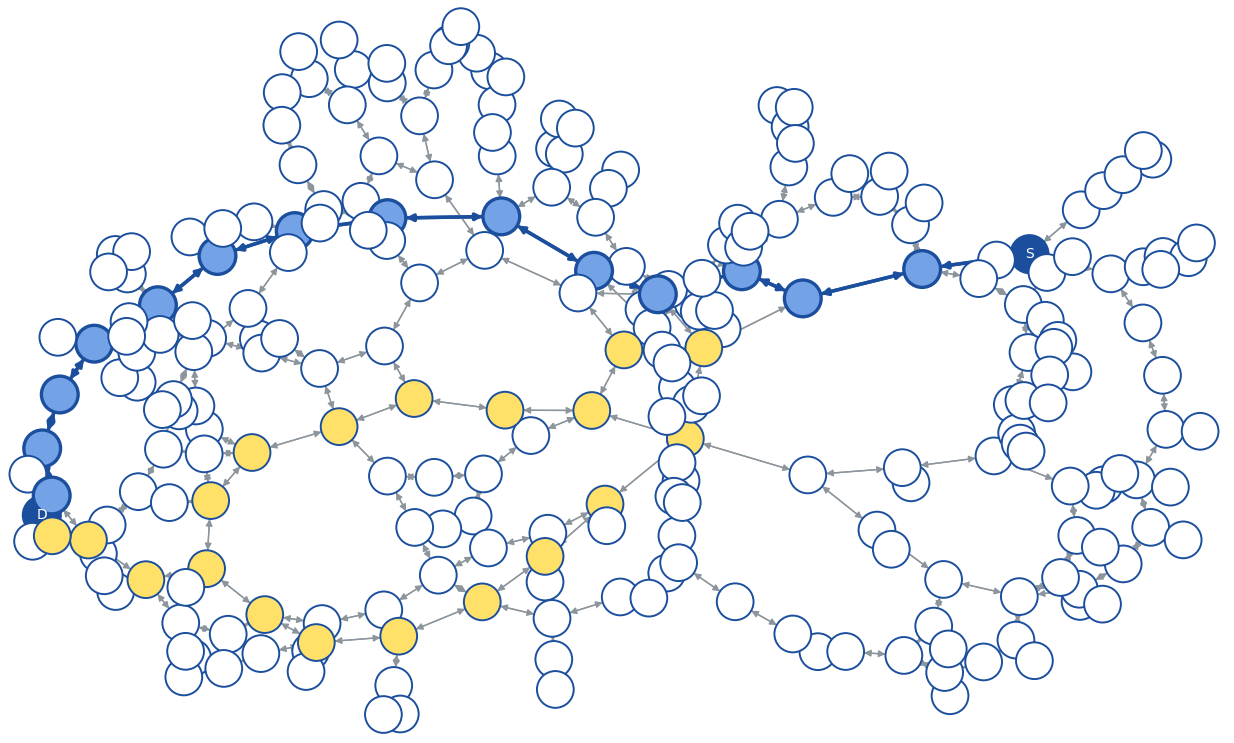}
\caption{Standard A* - 250 nodes}
\end{subfigure}
\hfill
\begin{subfigure}[t]{0.32\textwidth}
\includegraphics[width=\linewidth]{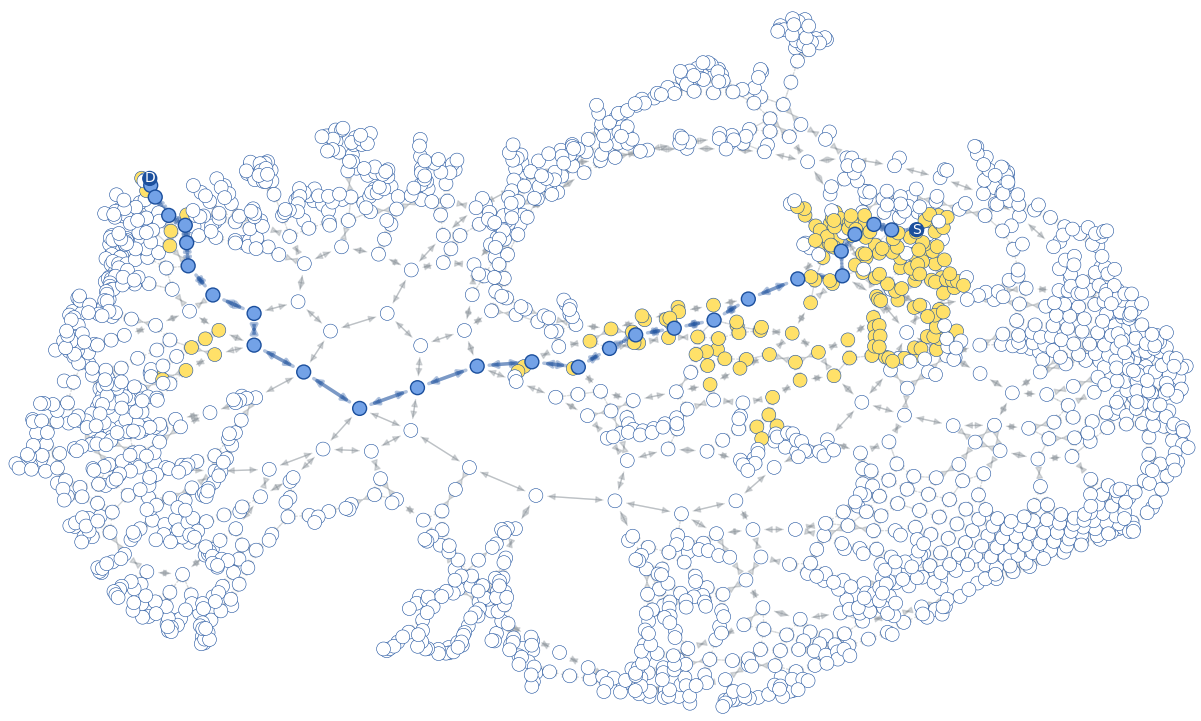}
\caption{Standard A* - 2000 nodes}
\end{subfigure}
 
\vspace{-0.20em}
 
% -------- Bottom row: LLM A* --------
\begin{subfigure}[t]{0.32\textwidth}
\includegraphics[width=\linewidth]{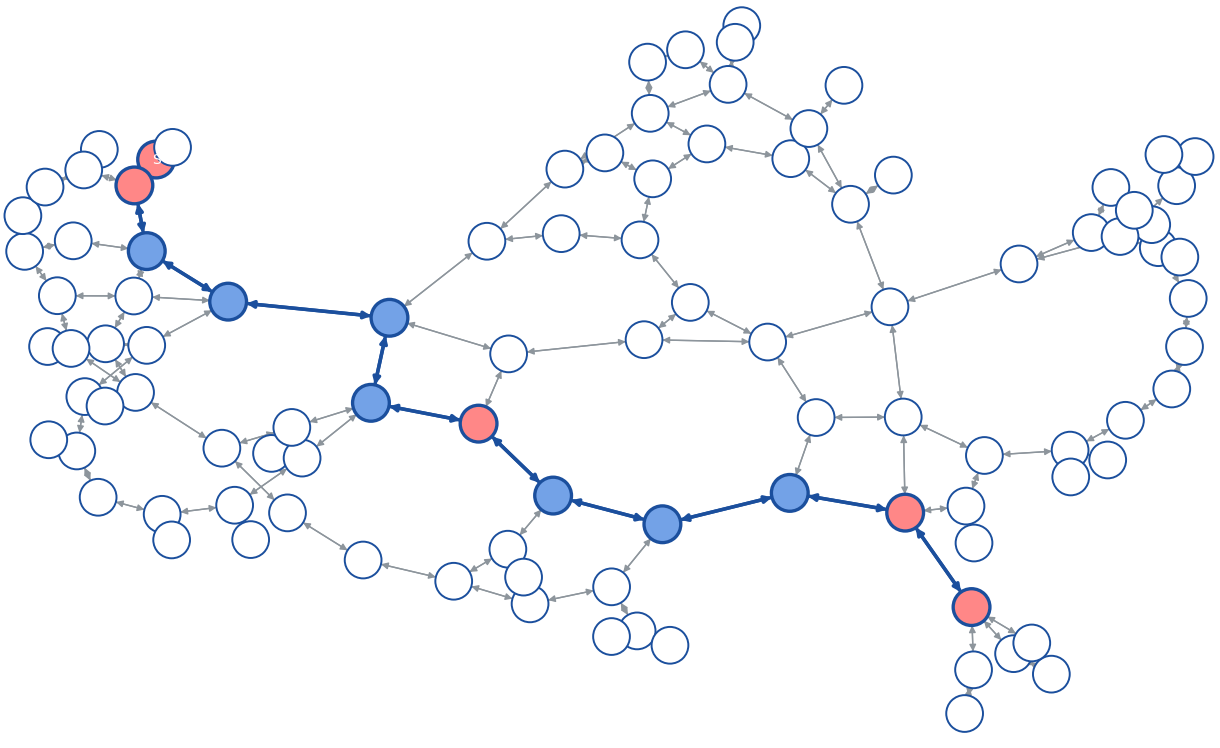}
\caption{LLM-aided A* - 100 nodes}
\end{subfigure}
\hfill
\begin{subfigure}[t]{0.32\textwidth}
\includegraphics[width=\linewidth]{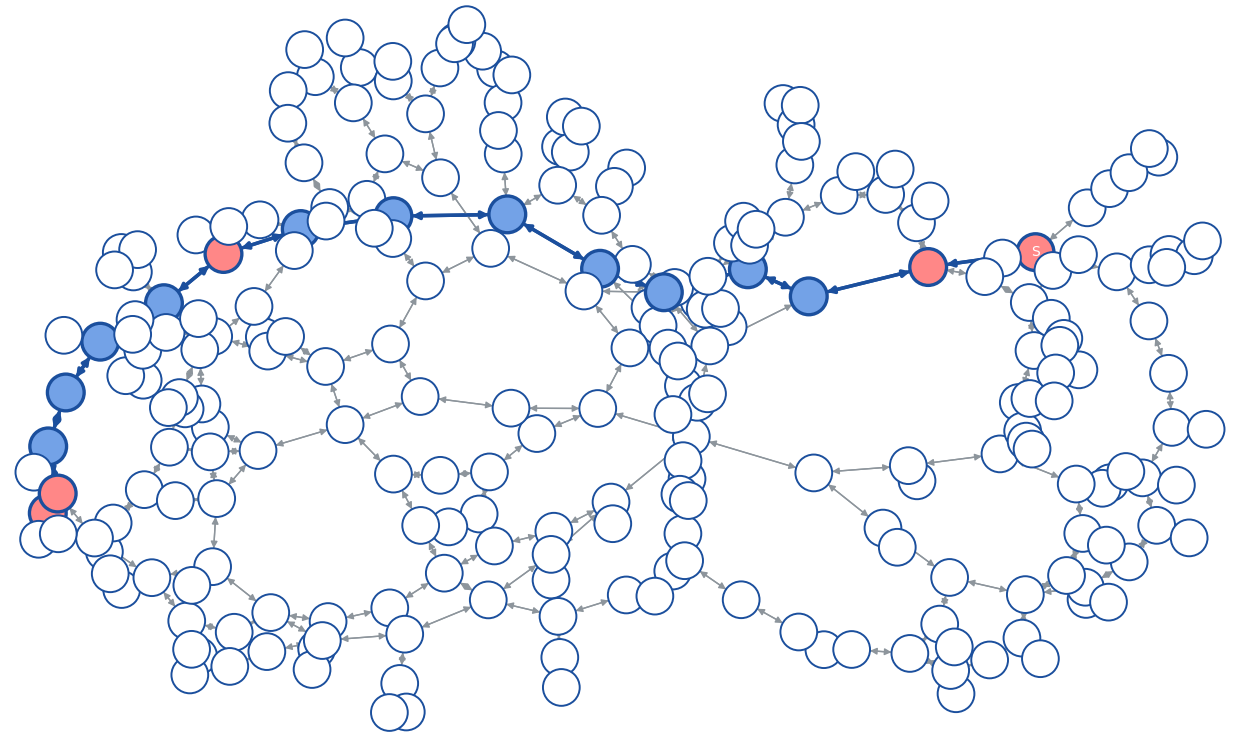}
\caption{LLM-aided A* - 250 nodes}
\label{fig:250_A*_LLM_final}
\end{subfigure}
\hfill
\begin{subfigure}[t]{0.32\textwidth}
\includegraphics[width=\linewidth]{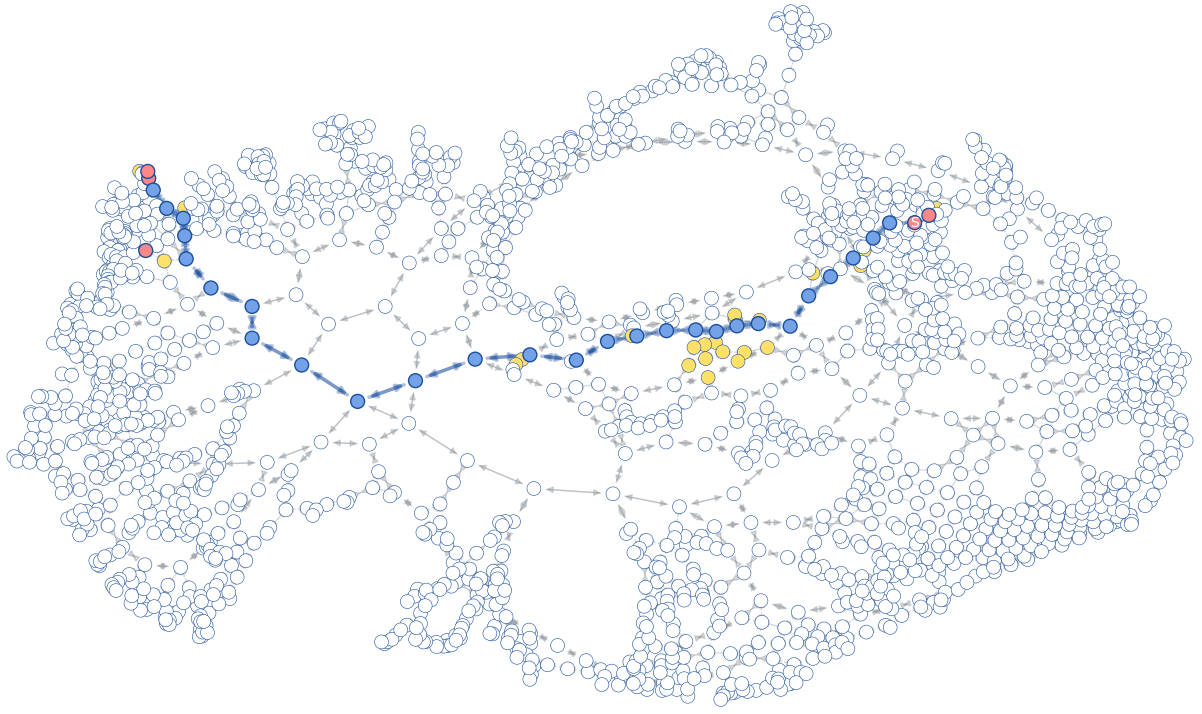}
\caption{LLM-aided A* - 2000 nodes}
\end{subfigure}
 
\caption{
Comparison of standard A* (top row) and LLM-aided A* (bottom row).
Yellow nodes represent expanded nodes. Dark blue nodes indicate the final path, while red nodes denote LLM-generated waypoints, which guide the search and reduce exploration.}
\label{fig:search_50-100_final}
\end{figure*}

% ============================
% ============================ Qualitative analysis — general success cases first, then the anomaly
Beyond the quantitative results reported in Table~\ref{fig:results_1}, we examine individual search instances to gain qualitative insight into the \ac{LLM}'s waypoint behavior.
As seen in Fig.~\ref{fig:search_50-100_final}, the LLM-aided A* consistently explores fewer nodes. The final paths (darker blue) appear comparable in quality between the two methods, suggesting the \ac{LLM} guidance doesn't sacrifice path optimality while significantly pruning the search space.
It can also be observed that the \ac{LLM}-generated waypoints at times focus the search on one of the possible path branches, Fig.~\ref{fig:250_A*_LLM_final}. This highlights their potential to disambiguate competing path branches, possibly guided by qualitative structural features of the graph that are not captured by the numerical heuristic alone.
 
\begin{figure}[t]
  \centering
  \begin{subfigure}[b]{0.45\textwidth}
    \centering
    \includegraphics[width=\linewidth]{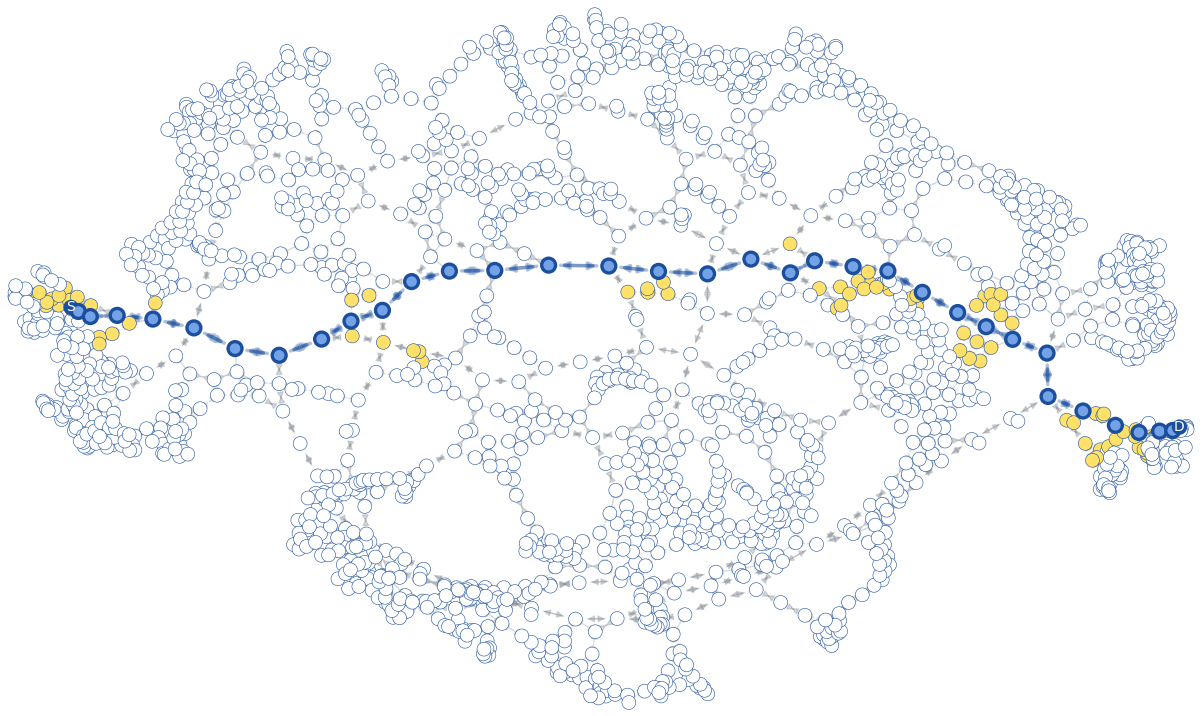}
    \caption{Standard A* - 2000 nodes}
    \label{fig:outlier}
  \end{subfigure}
  \hfill
  \begin{subfigure}[b]{0.45\textwidth}
    \centering
    \includegraphics[width=\linewidth]{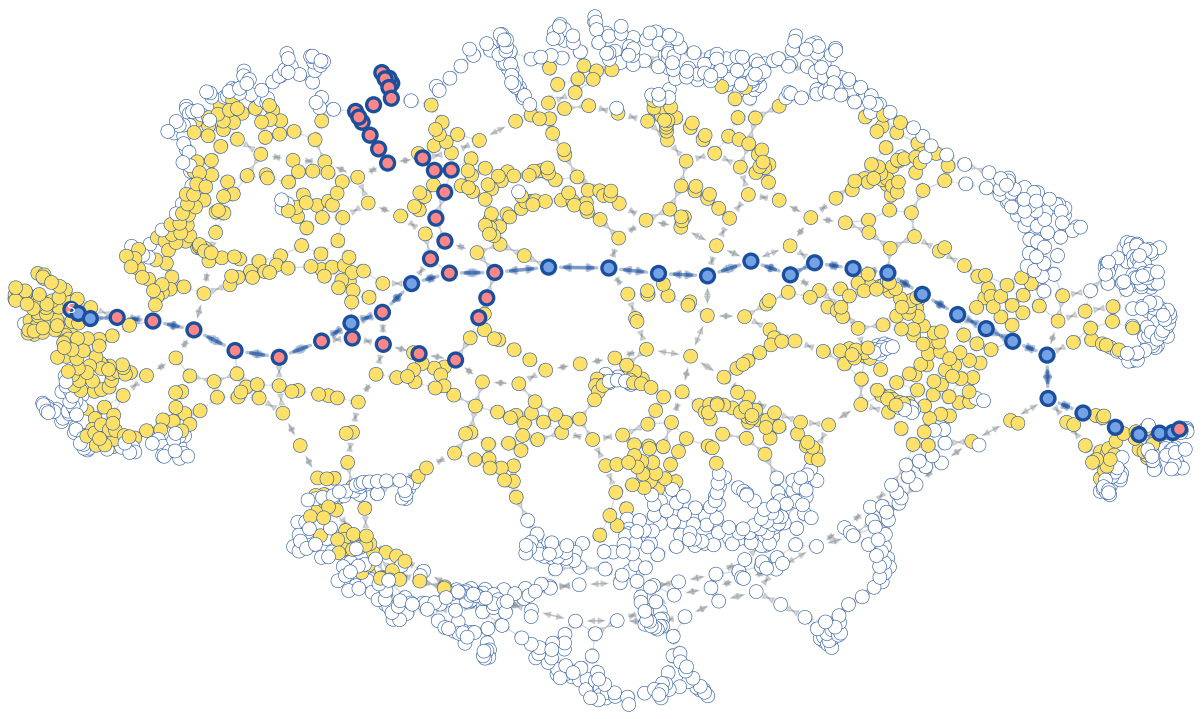}
    \caption{LLM-aided A* - 2000 nodes}
    \label{fig:outlier_LLM}
  \end{subfigure}
  \caption{Instance of an outlier case on a 2000-node graph. Yellow nodes represent expanded nodes. Dark blue nodes indicate the final path, while red nodes denote LLM-generated waypoints}
  \label{fig:outlier_v1}
\end{figure}

Having established that the guidance helps in the typical case, we now consider its worst-case behavior. Even in anomaly cases where the \ac{LLM}-generated waypoints deviate significantly from the optimal path, the A* algorithm still explores neighbors according to its cost function and heuristic, ultimately converging to a near-optimal solution. The effect of poor waypoints is therefore limited to increased exploration overhead rather than degraded path quality.
This behavior is illustrated in Fig.~\ref{fig:outlier_v1}, which depicts an outlier case where the LLM-generated waypoints misguide the search on a 2000-node graph. This interplay between \ac{LLM}-based guidance and algorithmic optimization provides
a degree of robustness to potential \ac{LLM} hallucinations. Since the \ac{LLM}'s output is used to bias the search priority rather than to dictate the solution, poorly generated waypoints primarily affect exploration efficiency rather than the quality of the final path.
 
\section{Conclusions And Future Work}\label{section_conclusions}
 
In this work, we introduce an LLM-aided A* algorithm for network graphs. By prompting an \ac{LLM} to generate intermediate waypoints, the proposed approach biases the A* expansion toward promising regions of the graph, reducing the number of explored nodes by around $50\%$ on network graphs with up to 2000 nodes while incurring only modest path cost increases in the range of 0.34--0.66 compared to the optimal path. Furthermore, we show that for graph-based tasks, enriching the \ac{LLM} input with compact, task-relevant structural features such as heuristic estimates is more effective than advanced prompting strategies; the latter help mainly when such structural information is absent and yield little additional benefit once it is provided. We also observe that the number of waypoints plays a critical role, with excessive waypoints over-constraining the search and diminishing performance gains. These findings highlight the potential of combining LLM-based global guidance with classical search algorithms, enabling more efficient exploration without sacrificing solution quality. Importantly, the \ac{LLM} acts as a guide rather than a solver, preserving the robustness of the underlying algorithm.
 
Several directions remain for future work. While the present study targets unweighted graphs to isolate the effect of \ac{LLM} guidance, an extended version will address weighted, resource-constrained shortest path problems. We plan to consider service function chain routing, where the \ac{LLM}'s ability to incorporate operator intent and policies expressed in natural language becomes a further asset. We also intend to formally bound the suboptimality introduced by the waypoint term, by deriving a bound on the worst-case excess over the optimal path cost to corroborate the empirical results reported here.

\bibliographystyle{IEEEtran}
\bibliography{ref}

\end{document}